\documentclass[a4paper]{siamltex}
\usepackage{amssymb}
\usepackage{amsbsy}
\usepackage{amsfonts}
\usepackage{amsmath}
\usepackage{amsopn}

\DeclareMathOperator{\re}{\mathbb{R}}
\DeclareMathOperator{\p}{\mathbb{P}}
\DeclareMathOperator{\e}{\mathbb{E}}

\DeclareMathOperator{\F}{\mathcal{F}}
\DeclareMathOperator{\D}{\mathbb{R}^\mathnormal{d}}
\DeclareMathOperator{\G}{\mathcal{G}}

\DeclareMathOperator{\T}{\mathrm T}
\def\S{\mathcal S}
\def\i{\infty}
\def\te#1{\mathrm{e}^{-r#1}}
\def\tld{\tilde}
\def\ovl{\overline}
\def\td{\text{\rm d}}

\begin{document}
\author{Mark Davis\thanks{Department of Mathematics, Imperial College London, SW7 2AZ London, UK\newline\hspace*{0.5cm}\texttt{mark.davis@imperial.ac.uk, j.obloj@imperial.ac.uk}}
 \and Jan Ob\L \'oj$^*$\thanks{Corresponding author. Research supported by a Marie Curie Intra-European Fellowship within the $6^{th}$ European Community Framework Programme
}}
\date{}
\title{Market completion using options}
\maketitle
\begin{abstract}
Mathematical models for financial asset prices which include, for example,
stochastic volatility or jumps are incomplete in that derivative securities
are generally not replicable by trading in the underlying. In earlier work
[\emph{Proc.~R.~Soc.~London}, 2004], the first author provided a geometric condition under which trading in the underlying and a finite number of vanilla
options completes the market. 
We complement this result in several ways.
First, we show that the geometric condition is not necessary and a weaker,
 necessary and sufficient, condition is presented. While this condition is
 generally not directly verifiable, we show that it simplifies to matrix non-degeneracy in a single point when the pricing functions are real analytic functions. In particular, any stochastic volatility model is then completed with an arbitrary European type option.
Further, we show that adding path-dependent options
such as a variance swap to the set of primary assets, instead of plain vanilla options, also completes the market. 
\end{abstract}



\section{Introduction} It is well known that the Black-Scholes financial
market model, consisting of a log-normal asset price diffusion and a non-random
money market account, is complete: every contingent claim is replicated
by a portfolio formed by dynamic trading in the two assets. Ultimately this
result rests on the martingale representation property of Brownian motion.
As soon as we attempt to correct the empirical deficiencies of the asset
model by including, say, stochastic volatility, completeness is lost if we
continue to regard the original two assets as the only tradables: there are
no longer enough assets to `span the market'. However there are traded options
markets for many assets such
as single stocks or stock indices, so it is a natural question to ask whether
the market becomes complete when these are included. An early result in this
direction was provided by Romano and Touzi \cite{RT97} who showed that a
single call option completes the market when there is stochastic volatility
driven by one extra Brownian motion (under some additional assumptions; see Section \ref{sec:csv} below). But providing a general theory has proved
surprisingly problematic. There are two main approaches, succinctly labelled
`martingale models' and `market models' by Schweizer and Wissel \cite{SchW06}.
In the former---which is the approach taken in Davis \cite{DAV04} and in this paper---one starts with a stochastic basis $(\Omega,\F,(\F_t)_{T\in \re_+},\p)$. $\p$ is a risk-neutral measure,
so all discounted asset prices are $\p$-martingales which can be constructed by conditional
expectation: the price process for an asset that has the integrable $\F_T$-measurable value
$H$ at some final time $T$ is $S^H_t=\e[e^{-r(T-t)}H|\F_t]$ for $t<T$, where $r$
is the riskless interest rate. The distinction
between an `underlying asset' and a `contingent claim' largely disappears
in this approach. A specific model is obtained by specifying some
process whose natural filtration is $(\F_t)$, for example a diffusion process
as in Section \ref{sec:setup} below. In a `market model' one specifies directly
the price processes of all traded assets, be they underlying assets or derivatives.
For the latter, say a call option with strike $K$ and exercise time $T$ on
an asset $S_t$, it is generally more convenient to model a proxy such as the
implied volatility $\hat{\sigma}_t$ which is related in a one-to-one way
to the price process
$A_t$ of the call by $A_t=\mathrm{BS}(S_t,K,r,\hat{\sigma}_t, T-t)$, where
$\mathrm{BS}(\cdots)$ is the Black-Scholes formula. This
is the approach pursued by Sch\"onbucher \cite{SCH99} and, in different
variants,  in recent papers
by Schweizer and Wissel \cite{SchW06} and Jacod and Protter \cite{JP06}.
This is not the place to debate the merits of these approaches; suffice it
to say that the problem with martingale models is that the modelling of
asset volatilities is too indirect, while the problem with market models
is the extremely awkward set of conditions required for absence of arbitrage.

The paper is organised as follows. We first describe our market, i.e.\ we model the factor process spanning the filtration and write assets prices as conditional expectations. Then in Section \ref{sec:stoch_crit} we give a necessary and sufficient condition for completeness of our market. Section \ref{sec:pde} explores the case when the coefficients of the SDE solved by the factor process and option payoffs are such that the prices are real analytic function of the time and the factor process. We show that the completeness question reduces from non-degeneracy of a certain matrix in the whole domain to its non-degeneracy in a single point. The result is then applied in Section \ref{sec:csv} to show completeness of stochastic volatility models. Section \ref{sec:assets} explores the use of path dependent derivatives, in particular variance swaps, in place of European type options and Section \ref{sec:concl} concludes.

\section{Market model}
\label{sec:setup}

Consider a market in which investors can trade in $d$ risky assets $A^1,\dots,A^d$ together with
a riskless money market account paying interest at a constant rate $r\geq 0$. We assume there is \emph{no arbitrage} in the market and we want to investigate \emph{market completeness} on $[0,T]$. We therefore assume existence of an equivalent martingale measure and we chose to work under this measure, which we denote $\p$. The market is spanned by some factor process. More precisely, market factors are modeled with a $d$-dimensional diffusion process $(\xi_t)_{t\geq 0}$, solution to an SDE:
\begin{equation}
\label{eq:factor}
\td\xi_t=m(t,\xi_t)\td t+\sigma(t,\xi_t)\td w_t,\quad \xi_0=x_0\in\D,
\end{equation}
where $w_t$ is a $d$-dimensional Brownian motion on $(\Omega,\F,\p)$, w.r.t.\ its natural filtration, and where we assume that 
\begin{equation*}
(A1) \quad
\begin{array}{l}
\text{$\sigma(t,x)\sigma(t,x)^{\T}$ is strictly positive definite for a.e.~$(t,x)\in(0,T)\times \D$,}\\
\text{$\eqref{eq:factor}$ has a unique strong solution.}
\end{array}
\end{equation*}
The assumption of ellipticity above seems natural and corresponds to \emph{full factor} representation.
The assumption that the state space of $\xi$ is the whole of $\D$ is a simplification which
allows us to expose the main ideas without superficial technicalities. In general one could take a general open connected set $\mathcal{D}\subset \re^d$ as the state space. Behaviour of $\xi$
at the boundaries would then imply the appropriate boundary conditions for PDE formulation in
\eqref{eq:PDE} below.

The semi-group of $(\xi_t)$
is denoted $(P_{u,t})$, i.e. $P_{u,t}h(x)=\e_{u,x}[h(\xi_t)]$, $u\leq t$, and $(\F_t)$ is the natural filtration of $(\xi_t)$, taken completed.

Traded assets are of European type, asset $A^i$ has a given payoff $h_i(\xi_{T_i})\geq 0$ at maturity $T_i$, larger
than the time-horizon on which we investigate market completeness, $T\leq T_i$. 
We assume implicitly $\e |h_i(\xi_{T_i})|<\infty$. As we choose to work under the risk neutral measure, the discounted
price process of an asset is a martingale. More precisely,
\begin{equation}
\label{eq:asset_price}
A^i_t=\e\Big[\te{(T_i-t)}h_i(\xi_{T_i})\Big|\F_t\Big],\quad 0\leq t\leq T_i.
\end{equation}
The setup we have in mind in particular is: $A^1_t=S^1_t=\exp(\xi^1_t)$ is the stock price itself,
however for the questions considered here there is no benefit from making this particular assumption. We could also consider assets with path-dependent payoffs, we will come back to this in Section \ref{sec:assets}.

The Markov property of $(\xi_t)$ implies that 
\begin{equation}
\label{eq:vi_def}
A^i_t=v_i(t,\xi_t),\quad\textrm{where }v_i(t,x)=\te{(T_i-t)}P_{t,T_i}h_i(x)
\end{equation}
\noindent and we assume that 
\begin{equation*}
(A2) \quad
\begin{array}{l}
\text{$v_i$ are of class $C^{1,2}$ on $(0,T)\times\D$, $1\leq i\leq d$.}\qquad\qquad\qquad\qquad\qquad\qquad
\end{array}
\end{equation*}
The latter property follows from the Feynman-Kac formula, or Kolmogorov's backward equation,
under mild regularity conditions (cf.~Friedman \cite[Chp.~6]{FRI75} or {\O}ksendal \cite[Chp.~8]{Oks03}). Let $G(t,x)$ be the matrix of partial derivatives,
\begin{equation}
\label{eq:matrixG}
G(t,x)=\Big(\frac{\partial v_i(t,x)}{\partial x_j}\Big)_{1\leq i,j\leq d}=\Big(\nabla
v_i(t,x)\Big)_{1\leq i\leq d}\; .
\end{equation}
Let $(M_t)$ be the martingale part of $(\xi_t)$, so that $\td M_t=\sigma(t,\xi_t)\td w_t$. Using the It\^o formula together with the fact that discounted prices are martingales, we see that
\begin{equation}
\label{eq:diffS}
\td \tld A_t=\td (\te{t}A_t)=\te{t}G(t,\xi_t)\td M_t=\te{t}G(t,\xi_t)\sigma(t,\xi_t)\td w_t,\quad t\leq T,
\end{equation}
where $A_t=(A^1_t,\dots,A^d_t)^{\T}$ is a column vector. In what follows, we refer to the above setup simply as \emph{the market}.

\section{Market completeness: stochastic criterion}\label{sec:stoch_crit}
A predictable process $(\alpha_t)$ in $\re^d$ with \[\e \int_0^T(\alpha^i_t)^2\td t<\infty,\quad1\leq i\leq d\] is a (general) \emph{trading strategy}, where $\alpha^i_t$ represents the number of units of $i^{th}$ asset invested at time $t$. It induces a self-financing portfolio $(X^\alpha_t)$ whose value behaves according to
\begin{equation}\label{eq:selffin}
\begin{split}
\td X^\alpha_t &=\sum_{i=1}^d \alpha^i_t\td A^i_t+\left(X^\alpha_t-\sum_{i=1}^d \alpha^i_t A^i_t\right) r\td t,\\
\td \tld X^\alpha_t &=\td (\te{t}X_t)=\sum_{i=1}^d \alpha^i_t\td \tld A^i_t\; .
\end{split}
\end{equation}
An $\F_T$-measurable claim can be hedged if it is a final value of a self-financing portfolio with some starting capital, that is if there exists $(\alpha_t)$ with 
\begin{equation}
H=\e H+X^\alpha_T=\e H + \mathrm{e}^{rT}\sum_{i=1}^d\int_0^T\alpha_t^i\td \tld A^i_t,\; a.s.
\end{equation}
We say that the market is complete if any claim can be hedged. More precisely, using the above, we make the following definition (where we absorb the factor $\mathrm{e}^{rT}$ in $\alpha$)
\begin{definition}\label{def:complete}
We say that the market on $[0,T]$ is complete if for any $\F_T$-measurable random variable $H$,
$\e H^2<\i$, there exists a predictable process $(\alpha_t)$ such that $\forall_{1\leq i\leq
d}$ $\e\int_0^T(\alpha_t^i)^2\td \langle
\tld A^i\rangle_t<\i$ and $H=\e H+\sum_{i=1}^d \int_0^T \alpha_t^i\td \tld A^i_t$.
\end{definition}

\noindent The assumption of integrability on $H$ is natural. General $H$ can still be represented but we need to authorize trading strategies $(\alpha_{t})$ such that $\int_0^T \alpha_t\td A_t$ is well defined while $\int_0^T \alpha^i_t\td A^i_t$ are \emph{not} well defined (see Jacod and Shiryaev \cite[Ex.~III.4.10]{JS03}) which makes little sense in market terms.

Let $\S$ be the set of zeros of the determinant of $G$ on $(0,T)\times\D$
\begin{equation}
\label{eq:Sdef}
\S:=\big(\det G\big)^{-1}\big(\{0\}\big)\subset (0,T)\times \D,
\end{equation}
which is a well defined Borel set. We can now state the characterization of market completeness.

\begin{theorem}
\label{thm:main}
Under the assumptions (A1) and (A2), the market is complete, in the sense of Definition \ref{def:complete}, if and only if 
$\int_0^T\mathbf{1}_{(t,\xi_t)\in\S}\td t=0$ a.s.
\end{theorem}

We can rephrase the above criterion by saying that $G(t,\xi_t)$ is non-singular $\td t$-a.e.\ on $(0,T)$ a.s. In particular, if the law of $\xi_t$ admits a density, the market is complete if $\S$ is of $(d+1)$-dimensional Lebesgue measure zero. The statement becomes "if and only if"
when the density is a.e.~strictly positive.
\proof ``$\Longleftarrow$"\\
Let $\ovl G^{-1}(t,x)=G^{-1}(t,x)\mathbf{1}_{(t,x)\notin\S}$ and let $H$ be any $\F_T$-measurable
random variable with $\e H^2<\i$. Using the representation theorem (cf.~Rogers and Williams \cite[V.25.1]{RW00}) for
$(\xi_t)$ we know there exists a predictable process $(\chi_t)$, $\e \int_0^T |\chi_t\sigma(t,\xi_t)|^2\td t<\i$, where $|x|^2=x^{\T}x$, with $H=\e H+\int_0^T \chi_t\sigma(t,\xi_t)\td w_t$. Put $\alpha_t:=\mathrm{e}^{rt}\chi_t\ovl G^{-1}(t,\xi_t)$ which is a predictable
process with 
$$\e\sum_{i=1}^d\int_0^T (\alpha^i_t)^2\td\langle\tld A^i\rangle_t\leq \e\int_0^T|\alpha_t G(t,\xi_t)\sigma(t,\xi_t)|^2\td
t\leq \e \int_0^T |\chi_t\sigma(t,\xi_t)|^2\td t<\i.$$
We have
$$\int_0^T\alpha_t\td \tld A_t=\int_0^T\chi_t\sigma(t,\xi_t)dw_t-\int_0^T\chi_t\sigma(t,\xi_t)\mathbf{1}_{(t,\xi_t)\in\S}\td
w_t=H-\e H,$$
where we used the assumption of the theorem (and Fatou lemma) to deduce that \\
$\int_0^T|\chi_t \sigma(t,\xi_t)|^2\mathbf{1}_{(t,\xi_t)\in\S}\td
t=0$ a.s. and thus $\int_0^T\chi_t\sigma(t,\xi_t)\mathbf{1}_{(t,\xi_t)\in\S}\td
w_t=0$ a.s.\smallskip\\
``$\Longrightarrow$"\\
Suppose that $\p\Big(\int_0^T\mathbf{1}_{(t,\xi_t)\in\S}\td t>0\Big)>0$.
Using Lemma \ref{lem:mss} in the Appendix choose a measurable function $\beta:(0,T)\times\D\to\re^d$ such that
\begin{equation}\label{eq:beta}
\left\{ \begin{array}{ll}
\beta(t,x)=0 & \textrm{for }(t,x)\notin\S\\
 G(t,x)\sigma(t,x)\sigma(t,x)^{\T}\beta(t,x)^{\T}=0,\; |\beta(t,x)\sigma(t,x)|^2=1 & \textrm{for } (t,x)\in\S\\
\end{array} \right. 
\end{equation}
and let $H=\int_0^T\beta(t,\xi_t)\sigma(t,\xi_t)\td w_t$. Naturally $H$ is $\F_T$-measurable and
\[\e H^2=\e\int_0^T|\beta(t,\xi_t)\sigma(t,\xi_t)|^2\td t=\e \Big[\int_0^T\mathbf{1}_{(t,\xi_t)\in\S}\td t\Big] \in (0,T].\] For any predictable process $(\alpha_t)$
with $\e\int_0^T|\alpha_t G(t,\xi_t)\sigma(t,\xi_t)|^2\td t<\i$ we have
$$\e\big[H\cdot\int_0^T\alpha_t\td\tld A_t\big]=\e\Big[\int_0^T\te{t}\alpha_t G(t,\xi_t)\sigma(t,\xi_t)\sigma(t,\xi_t)^{\T}\beta(t,\xi_t)^{\T}\td
t\Big]=0,$$
which proves that $H$ is orthogonal to the space generated by the stochastic integrals w.r.t.
$\tld A$ and the market is incomplete.
\hfill$\square$

\section{Market completeness: PDE approach}
\label{sec:pde}
In Theorem \ref{thm:main} we stated a general necessary and sufficient condition for our market
to be complete. So far however, we did not provide any easy means to verify that the condition holds.
This is the purpose of this section. We exploit the Feynman-Kac formula to rephrase our condition
in terms of PDEs and then use interior regularity of solutions of PDEs.

Let $\G_t$ be the generator of $(P_{t,t+u})$ acting on regular functions $f:\re^d\to\re$ via
\begin{equation}
\G_t f(x)=\nabla f(x)m(t,x)+\frac{1}{2}\sum_{i,j=1}^d\big(\sigma(t,x)\sigma(t,x)^{\T}\big)_{i,j}\frac{\partial^2 f}{\partial x_i\partial x_j}(x).
\end{equation}
The Feynman-Kac formula, provided we can justify its application, shows that the functions $v_i$ in \eqref{eq:vi_def} satisfy
\begin{equation}
\begin{split}\label{eq:PDE}
\ovl\G v_i:=\frac{\partial v_i}{\partial t}+\G_t v_i-rv_i &=0,\quad (t,x)\in (0,T_i)\times\D\\
v_i(T_i,x)&=h_i(x),\quad x\in\D.
\end{split}
\end{equation}
We need the above for our next result, so we assume explicitly that it holds:
\begin{equation*}
(A3) \quad
\begin{array}{l}
\text{$m,\sigma,h_1,\dots,h_d$ are such that $\xi_t$ admits a.e.~positive density on $\D$, $t\leq T$, and} \\ 
%\text{$h_i$ have at most polynomial growth,}\\
\text{$v_i\in C^{1,2}$ are the unique (under suitable growth conditions) solutions of \eqref{eq:PDE}.}
\end{array}
\end{equation*}
One may choose various sets of conditions on $m$ and $\sigma$ and on the growth of $v_i$ and on the payoffs which grants $(A3)$, we refer the reader to Friedman \cite[Sec.~6.4-6.5]{FRI75} for examples.

Theorem \ref{thm:main}, in the above setting, states that the market is complete if and if only $\mathcal{S}$ is of zero Lebesgue measure. Thus, we can rephrase the question of \emph{market completeness} into an equivalent question about properties of a solution to a system of PDEs. This seems to be a hard question in general, but we can solve it neatly in the case when $v_i$
are real analytic. The idea of exploiting properties of analytic functions to understand market
completeness goes back to Bj\"ork, Kabanov and Runggaldier \cite{BKR97}\footnote{We want to thank
Sara Biagini for bringing this to our attention.}. Naturally there the authors
choose the dynamics of the forward rate, which in our setting would be equivalent to \emph{taking}
entries of the matrix $G$ to be real analytic, while here it is something we have to derive.

\begin{theorem}\label{thm:an_sol}
Suppose that $(A1)$ and $(A3)$ hold and that further $m,\sigma,h_i$ are such that $v_i:(0,T)\times \D\to\re$, $1\leq i\leq d$, are real analytic functions. Then the market is complete if and only if there exists a point $(t_0,x_0)\in (0,T)\times \D$  such that $G(t_0,x_0)$ is non-singular.
\end{theorem}
\proof
Where $v_i$ are analytic so are their partial derivatives, which form the entries of the matrix $G$. Products and sums of analytic functions are also analytic and thus $\det G:(0,T)\times\D\to\re$ is real analytic. In consequence, it is either equal to zero on the whole domain of analyticity or its set of zeros $\mathcal{S}$ is Lebesgue negligible (cf.~Krantz and Parks \cite[p.83]{KP02}). By Theorem \ref{thm:main}, and since $\xi_t$ admits a density, the market is complete if and only if we are in the latter case. This in turn is equivalent to non-degeneracy of $G$ at least at one point $(t_0,x_0)$ since then, by continuity of $\det G$, $G$ is non-degenerate in some neighbourhood of $(t_0,x_0)$ of positive measure.
\hfill$\square$

The natural question resulting from the Theorem is: when are the solutions to the parabolic equation
\eqref{eq:PDE} real analytic? It turns out that in general, assuming the coefficients $m,\sigma$ are real analytic, the solutions $v_i$ are not jointly analytic in $(t,x)$ but are 
 are analytic in space and of Gevrey class $2$ in time (cf.~H\"ormander \cite[Sec.~11.4]{HOR83}
 and Eidelman \cite[Thm.~II.7.2]{Eid69}). Roughly speaking, the existence of non-analytic solutions is linked to the possibility of very rapid growth. It appears one can guarantee analyticity of $v_i$'s when they remain bounded, for this we need\footnote{Results referred to in $(A4)$ are stated in \cite{Tetal96} without a proof as a straightforward generalisation of the main theorems therein. However after a discussion with the authors it appears the proof might be more involved then they suggest.}:
\begin{equation*}
(A4) \quad
\text{The results stated in Remark $4.1$ in Tak\'a\v c \emph{et al.} \cite[Sec.~4]{Tetal96} hold
true.
}
\end{equation*}
\begin{corollary}\label{cor:bounded_analytic}
Suppose that $(A1)$, $(A3)$ and $(A4)$ hold and that $m,\sigma$ are real analytic with \begin{equation}
\label{eq:sell}
\sum_{i,j=1}^d \big(\sigma(t,x)\sigma(t,x)^{\T}\big)_{i,j} q_iq_j\geq \mathrm{c}|q|^2,\quad\forall q\in\re^d,(x,t)\in\D\times(0,T),
\end{equation}
for some $c>0$. If $h_i$, $1\leq i\leq d$, are bounded then $v_i:(0,T)\times \D\to\re$, $1\leq i\leq d$, are real analytic functions. In particular, the market is then complete if and only if there exists a point $(t_0,x_0)\in (0,T)\times \D$  such that $G(t_0,x_0)$ is non-singular.
\end{corollary}
\\Using the stochastic representation in \eqref{eq:asset_price}-\eqref{eq:vi_def} we have $v_i(t,x)=\mathrm{e}^{-r(T_i-t)}\e_{t,x}[h_i(\xi_{T_i})]$
so it is clear that $v_i$ are bounded when $h_i$ are bounded. 
%This can also be argued on the grounds of PDE theory via maximum principle for parabolic operators.
Analyticity of $v_i$ then follows from Tak\'a\v c \emph{et al.} \cite[Sec.~4]{Tetal96} and $(A4)$
looking at the time-reversed (adjoint) equation.

The Corollary is useful since bounded payoffs are naturally encountered in financial market, a prime example being put options. Furthermore, from analyticity of prices of put options we
can deduce analyticity of prices of call options via call-put parity. We will exploit this when examining stochastic volatility models in Proposition \ref{prop:svol} below. 

However, the condition that the payoffs are bounded seems to be unnatural in our setup. We feel that to exclude the possibility of rapid growth of $v_i$ it should be sufficient to make assumptions
about the integrability of $h_i(\xi_T)$. Much to our surprise, we have not found any works discussing it. The question of time-space analyticity of solutions to parabolic PDEs with terminal condition seems to be an open and, in the
light of our present study, interesting question. However, this is a PDE theoretical
question which goes well beyond the scope of this paper.
 
\subsection{Example: correlated Brownian motion}\label{sec:ex}
We propose to look at the simple example of Brownian motion (i.e.~of the heat equation).
It is known that the heat equation admits non-analytic solutions -- this was already
discussed by Holmgren \cite{HOL08} (see Bilodeau \cite{BIL82}). However, these can be excluded by imposing growth restrictions on the solutions.

Consider $\xi_t=\sigma w_t$ a correlated Brownian motion and assume $h_i$ have at most polynomial growth. In this simple case we know the semi-group so that we can write function $v_i$ explicitly.  Assume for simplicity that all options mature at time $T$ and set $r=0$. 
If we set $\tilde{v}_i(t,x)=\e_x \tilde{h}_i(w_{T-t})$ with $\tilde{h}_i(x)=h_i(\sigma x)$ then $v_i(t,x)=\tilde{v}_i(t,\sigma^{-1}x)$. Therefore, with no loss of generality we can take $\sigma=Id$,
the identity matrix. We then have
\begin{equation}\label{eq:vforBM}
v_i(t,x)=\e_x[h_i(\xi_{T-t})]=\frac{1}{(2\pi(T-t))^{d/2}}\int_{\re^d} h_i(y)\mathrm{e}^{-\frac{|u-x|^2}{2(T-t)}}\td
u.
\end{equation}
We can extend $v_i$ via \eqref{eq:vforBM} to all points 
\begin{equation}\label{eq:complex_set_def}
(\mathbf{t},\mathbf{x})\in \left\{ (t+\mathrm{i}\tau,x+\mathrm{i}y): 0<t<T, x\in \D, |\tau|<T-t,|y|<
\sqrt{T-t}\right\}\subset \mathbb{C}^{d+1}.
\end{equation}
Indeed, for $(\mathbf{t},\mathbf{x})$ as in \eqref{eq:complex_set_def} one can verify that 
\begin{equation*}
\left|\exp\left(-\frac{|u-\mathbf{x}|^2}{2\mathbf{t}}\right)\right|\leq C\exp\left(-\delta\frac{|u-x|^2}{2t}\right),\quad
u\in\D,
\end{equation*}
for some universal constants $C,\delta$. Differentiating under the integral, using the growth restriction on $h_i$, we see that $v_i(\mathbf{t},\mathbf{x})$ is continuously differentiable, and thus analytic, with respect to $\mathbf{t}$ and all $\mathbf{x}_i$, $1\leq i\leq d$ (see John \cite[Sec.~III.10]{JOH71} for a one-dimensional version of the argument). Application of Hartogs' theorem (cf.~Cartan \cite[IV.5.2]{CARTAN63}) yields joint analyticity of $v_i$ in $(\mathbf{t},\mathbf{x})$
and in particular real analyticity on $(0,T)\times\D$.

We observe that alternatively one can obtain analyticity of $v_i$ by direct, albeit tedious, computation. In fact
\eqref{eq:vforBM} implies that $v_i$ have polynomial growth in $(T-t)x$. More precisely, if $h_i(x)\leq a |x|^{2k}$ then we have $v_i(t,x)\leq 4^ka\e[|w_{1}|^{2k}]\cdot (T-t)^{k}|x|^{2k}$. One can then differentiate under the integral in \eqref{eq:vforBM} and obtain analogous bounds for all the derivatives of $v_i$ which then implies $v_i$ is real analytic (cf.~Krantz and Parks \cite[Prop.~2.2.10]{KP02}). 
We note again that it is not true that any solution to the heat equation
is real analytic in $(t,x)$ and here we rely strongly on the growth assumptions of $v_i(T,x)=h_i(x)$.
However it is clear that the assumption of polynomial growth of $h_i$ could be weakened to, say,
$h_i(x)\leq \alpha_1\exp(\alpha_2 |x|)$.

We want to stress that the condition of non-degeneracy at least at one point is important in
Theorem \ref{thm:an_sol}. As a counterexample, consider the situation when our assets are
call options with different strikes. More generally, take $\sigma$ to be non trivial and suppose that the payoffs depend only on the stock, i.e. $h_i(x)=h_i(x_1)$. We can represent $\xi_{T-t}^j=\tld q_j\xi^1_{T-t}+\tld c_j N$, with $N$ independent of $\xi^1$ which gives 
\begin{equation}
\label{eq:onefactordep}
\e_0[h_i(x+\xi^1_{T-t})\xi^j_{T-t}]=\tld q_j\e_0[h_i(x+\xi^1_{T-t})\xi^1_{T-t}].
\end{equation}
Working out the derivatives matrix we get
\begin{equation}
G(t,x)=\frac{1}{T-t}\Big(\e_x[ h_i(\xi_{T-t})(\xi^j_{T-t}-x_j)]\Big)_{i,j\leq d}\cdot Q,
\end{equation}
where $Q=(\sigma^{-1})^{\T}\sigma^{-1}$, and \eqref{eq:onefactordep} readily implies that $\det G\equiv 0$. In fact, in this setup whenever the payoffs depend only on $(d-2)$ or fewer factors  $G$ is degenerate and market is incomplete.
This is still true even if we consider options with different maturities.

A simple example when the market is complete is obtained taking $\sigma=Id$ and $h_i(x)=x_i^2$. Then $G$ is a diagonal matrix with $G(t,x)_{ij}=2x_i\mathbf{1}_{i=j}$. The set of singularities  
$\mathcal{S}=(0,T)\times \{x: x_1\cdot\ldots\cdot x_d=0\}$ has $(d+1)$-dimensional Lebesgue measure zero and the market is complete by Theorem \ref{thm:main}.

\section{Complete stochastic volatility models}\label{sec:csv}

We specialize now to the case $d=2$ which corresponds to stochastic volatility models. 
We use the conventional notation so that $A_t^1=S_t=\exp(\xi^1_t)$ is the stock price process
and $\xi_t^2=Y_t$ is the process driving the volatility. The process $(S_t,Y_t)$ under the risk-neutral
measure $\p$ satisfies
\begin{equation}\label{eq:svol_sde}
\left\{ \begin{array}{ll}
\td S_t=rS_t\td t+\sigma(t,S_t,Y_t)S_t\td w^1_t, & S_0=s_0>0,\\
\td Y_t=\eta(t,S_t,Y_t)\td t + \gamma(t,S_t,Y_t)\td \tld w_t, & Y_0=y_0,
\end{array} \right. 
\end{equation}
where $\tld w_t=\rho(t,S_t,Y_t) w^1_t+\sqrt{1-\rho(t,S_t,Y_t)^2}\,w^2_t$. We assume the coefficients
are such that $(S_t,Y_t)$ is well defined as the unique strong solution of \eqref{eq:svol_sde}
with $S_t>0$. The process $\xi^1_t=\log(S_t)$ is then well defined and It\^o's formula shows
it satisfies 
$$\td \xi^1_t=\big(r-\frac{\sigma(t,\exp(\xi^1_t),\xi^2_t)^2}{2}\big)\td t+\sigma(t,\exp(\xi^1_t),\xi^2_t)\td
w^1_t,$$
so that $\xi=(\xi^1,\xi^2)$ solves equation of the type \eqref{eq:factor}. Romano and Touzi \cite{RT97} were able to show that the above market is completed with a European
call under the additional assumptions that $\sigma, \eta, \gamma,\rho$ do not depend on $S_t$. We replace these assumptions with the analyticity assumption.
\begin{proposition}\label{prop:svol}
Consider assets $A^1_t=S_t$ and $A_t^2=v_2(t,S_t,Y_t)$ a European option with a payoff $h(S_T)\geq 0$, where $h$ is an arbitrary not-affine function. Under $(A1)$ and $(A3)$, if $v_2$ is real analytic then the market is complete.\\
In particular, if $m,\sigma$ are analytic and \eqref{eq:sell}, $(A4)$ hold then the market is completed by trading in a European call or put option.
\end{proposition}
\\We note that when $m,\sigma$ are real analytic and \eqref{eq:sell} holds then $(A1)$ and $(A2)$
amount to growth restrictions in $x$ on $m,\sigma$. When \eqref{eq:svol_sde} has a unique strong
solutions and $v_i$ are the unique solutions of \eqref{eq:PDE} then also $(\xi_t)$ admits density which is the fundamental solution of $\mathcal{G}_t+\partial/\partial t$ (cf.~Friedman \cite[Thm.~5.4]{FRI75}).
The density is then real analytic in the space variables $x$ and it follows
that it is a.e.~positive as it is non-negative and integrates to $1$.
\proof
We use $s$ for values of $S_t$ and $x$ for values of $\xi^1_t$ so that we always have $s=\exp(x)$
and we use $s,x$ interchangeably. We choose to write $v_2$ as a function of $S_t$, as this is
more natural, and transcribe the equation for $v_2$ into $(t,s,y)$ coordinates.\\
First note that if $\sigma$ does not depend on $Y$, i.e. $\forall t\leq T, s>0, y_1,y_2$, $\sigma(t,s,y_1)=\sigma(t,s,y_2)$,
than the market is complete just by trading in the stock (we have in fact a local volatility
model driven by one Brownian motion). We fix a non-affine payoff function $h\geq 0$
and we assume that $\sigma$ depends on $y$. Recall that $\sigma$ is real analytic and so are
its partial derivatives and the set of zeros of an analytic function is either equal to the whole
domain or is negligible. As $\sigma$ depends on $y$ we deduce that the set of zeros of $\frac{\partial \sigma}{\partial y}$ is negligible. More precisely, for a.e $(t,s)\in (0,T)\times\re_+$ the set of zeros of $\frac{\partial \sigma(t,s,y)}{\partial y}$ is negligible and in particular there exist $y_1,y_2$ in the support of $Y_t$, such that $\sigma(t,s,y_1)\neq \sigma(t,s,y_2)$.

As our first asset is the first factor, we have $v_1(t,s,y)=s=\exp(x)$ and the first row of $G(t,x,y)$ is simply $(\exp(x),0)$. By Theorem \ref{thm:an_sol} the market is complete if and only if $G$ is non-degenerate at least at one point, which is in turn equivalent to showing that there exist $(t,s,y)$ such that $\frac{\partial}{\partial y} v_2(t,s,y)\neq 0$. Suppose to the contrary that $v_2(t,s,y)=v_2(t,s)$ is independent of $y$. It follows from \eqref{eq:PDE}, written in
$(t,s,y)$ and not $(t,x,y)$ coordinates, that $v_2$ satisfies
\begin{equation}\label{eq:pde_svol_deg}
\frac{\partial v_2}{\partial t}+rs\frac{\partial v_2}{\partial s}+\frac{\sigma^2(t,s,y)s^2}{2}\frac{\partial^2v_2}{\partial
s^2}-rv_2=0.
\end{equation}
The only term in \eqref{eq:pde_svol_deg} which depends on $y$ is $\sigma$. The dependence of
$\sigma$ on $y$, as discussed above, then implies that $\frac{\partial^2 v_2}{\partial s^2}\equiv 0$ a.e.\ for $(t,s)\in (0,T)\times\re_+$, so that $v_2(t,s)$ is linear in $s$.
Writing $v_2(t,s)=\alpha(t)+s\beta(t)$
and plugging in \eqref{eq:pde_svol_deg} we see that $\beta'(t)=0$ and $\alpha'(t)=r\alpha(t)$.
It follows that $h$ is an affine function which gives the contradiction.\\
It follows from Corollary \ref{cor:bounded_analytic} and $(A4)$ that if $m,\sigma$ are analytic and \eqref{eq:sell} holds then $v_2$ is real analytic when $h(x)=(K-x)^+$ is a put payoff. Analyticity of the price
of a call option then follows from put-call parity (cf.~Karatzas and Shreve \cite[p.~50]{KS98}).
\hfill$\square$

\section{On the choice of assets completing the market}
\label{sec:assets}

We introduced in Section \ref{sec:setup} the general setup of market driven by $d$-dimensional
factor process $(\xi_t)$ in which we can trade in $d$ assets $(A_t^1,\dots,A^d_t)$. As we work under risk-neutral measure, assets prices are specified uniquely via the corresponding maturities $T_i$ and payoffs $h_i(x)$, $A^i_t=\e[\te{(T_i-t)}h_i(\xi_T)|\F_t]$, where we assume $T^i\geq T$.
In the basic setting $A^1=S=\exp(\xi^1)$ is the stock price itself and other option payoffs depend on the first coordinate only: $h_i(x)=h_i(x_1)$. More generally, we can think of having $n$ stocks, $A^i=S^i$, $1\leq i\leq n$. Other assets then could include some basket options with payoffs $h(x)=h(x_1,\dots,x_n)$. 
However so far we have allowed only European style options. In various markets
some path-dependent options, such as variance swaps, are liquid and it may be natural  to use them to complete the market. 
We show now how this can be incorporated in our setup.

Let $X_t=\log(S_t/S_0)$, where $S_t=A^1_t$ is the stock price process. The variance swap pays the quadratic variation $V_T=\langle X\rangle_T$ at maturity
$T$ (cf.~Gatheral \cite[Chp.~11]{G06}). The
process $\ovl X_t=X_t-rt=\log(\tld S_t/S_0)$ differs from $X$ by a finite variation term, so
that $\langle X\rangle_T=\langle \ovl X\rangle_T$. The price of a variance swap at time $t$ is given by $V_t=\e [\te{(T-t)}\langle X\rangle_T|\F_t]=v_V(t,\xi_t)$. 
The following derivation is well known
\begin{equation}
\label{eq:varswap}
\begin{split}
&\ovl X_T=\int_0^T\frac{\td \tld S_t}{\tld S_t}-\frac{1}{2}\int_0^T\frac{\td \langle \tld S\rangle_t}{\tld
S^2_t}=\int_0^T\frac{\td \tld S_t}{\tld S_t}-\frac{1}{2}\langle \ovl X\rangle_T,\quad \textrm{and
thus}\\
&V_t=2\te{(T-t)}\int_0^t\frac{\td \tld S_u}{\tld S_u}-2\e\Big[\te{(T-t)}\log(\tld S_T/S_0)\Big|\F_t\Big].
\end{split}
\end{equation}
 Suppose $d^{th}$ asset's payoff is given as $h_d(x)=\log(x_1/S_0)$. It follows from \eqref{eq:varswap} that
 \begin{equation}
 \label{eq:varswap2}
 \begin{split}
&\td \tld V_t=\td (\te{t}V_t)=2\te{T}\frac{\td \tld A^1_t}{\tld A^1_t}-2\td \tld A^d_t,\quad \textrm{or equivalently}\\
& \nabla v_V(t,x)=\frac{2\te{(T-t)}}{v_1(t,x)}\nabla v_1(t,x)-2\nabla v_d(t,x).
\end{split}\end{equation}
In consequence, the rank of a matrix $G$, whose first row is $\nabla v_1$, remains unchanged when we replace the row $\nabla v_d$ by $\nabla v_V$. We state this as a proposition.
\begin{proposition}
Consider a market model of Section \ref{sec:setup} with assets $A=(A^1_t,\dots,A^d_t)$, where $A^1=S$ is the stock price and $A^d$ has payoff $\log S_T$ at maturity $T$. Trading in $A$ completes the market if and only if trading in $\ovl A$ completes the market, where $\ovl A^i=A^i$, $i<d$ and $\ovl A^d_t=V_t$.
\end{proposition}

The method presented above allows to investigate other path-dependent options, as long as we can write them as sum of trades in the remaining assets plus a different asset with a European payoff. We could for example consider an option paying $\langle S\rangle_T$.

\section{Conclusions}\label{sec:concl}
To model realistically the dynamics of the stock price process one typically needs to consider models driven by more factors than just one Brownian motion. This naturally leads to market incompleteness
when only trading in the stock is considered. Pricing of derivatives is no longer unique. It
is in fact a challenging problem which has been extensively studied. However, in present markets one does
not need to price all derivatives. Indeed, some options are so liquid that they should be treated
as inputs of the model. This was the starting point of our work.

The basic rule of the thumb is naturally: in order to have a complete market take as many assets (including the stock itself) as you have random processes spanning the filtration (see for example Karatzas and Shreve \cite[Thm.~1.6.6]{KS98}). The question
is then: when is this intuition actually correct? Theorem \ref{thm:an_sol} shows that in the most regular case it is essentially always correct . More precisely, we consider market model written as an SDE with coefficients and payoffs such that pricing functions $v_i$ in \eqref{eq:vi_def}
are real analytic. Then it suffices to check that the matrix governing the evolution
of asset prices is non-singular in one point to deduce that the set of assets completes the
market. In particular, in Proposition \ref{prop:svol} we show that then a stochastic volatility
model is always completed by a single European option.

It seems, there are two main open questions resulting from the present work. An
analogue of Proposition \ref{prop:svol} for higher-dimensional models would complement Theorem
\ref{thm:an_sol} and provide a full understanding of market completeness with options. Throughout,
we considered the SDE \eqref{eq:factor} which is driven by a $d$-dimensional Brownian motion. The second remaining challenge is to extend this to the discontinuous setup. This is the subject of our further research.

Finally, in view of our results it is crucial to understand when a solution to \eqref{eq:PDE}
is real analytic and in particular to give a complete proof of $(A4)$. We hope this will motivate further study in the theory of PDEs.

\section{Appendix}
For completeness, we give the proof of the following measurable selection lemma used in the proof of Theorem \ref{thm:main}.
\begin{lemma}\label{lem:mss}
Let $G$ be defined via \eqref{eq:matrixG}. There exists a measurable function $\beta:(0,T)\times\D\to\re^d$
satisfying \eqref{eq:beta}.
\end{lemma}
\proof
Let $Z=(0,T)\times\D$ and define $\Gamma(z)=\Gamma(t,x)=G(t,x)\sigma(t,x)\sigma(t,x)^{\T}$, $z=(t,x)$. Matrix $\Gamma(z)$ induces a linear map on $\re^d$ and $\Gamma:Z\times\re^d \to \re^d$ given by $(z,y)\to \Gamma(z)y$ is a continuous function. For $z\in Z$ define
\begin{equation}\label{eq:functionalF}
F(z)=\left\{ \begin{array}{ll}
\{0\} & \textrm{if }\det \Gamma(z)\neq 0\\
Ker(\Gamma(z))\cap\{y:|y|^2=1\}& \textrm{otherwise.}\\
\end{array} \right. 
\end{equation}
$F(z)$ is a non-empty closed set in $\re^d$ for any $z\in Z$. Let $U\subset\re^d$ and put $F^-(U)=\{z\in Z: F(z)\cap U\neq \emptyset\}$. We will now argue that $F^-(U)$ is a measurable set for any closed
$U$. Let $\tld U=U\cap(\{y:|y|=1\}\cup \{0\})$ and observe that
\begin{equation}
F^-(U)=\mathrm{p}_z\left(\Gamma^{-1}(\{0\})\cap(Z\times\tld U)\right),
\end{equation}
where  $\mathrm{p}_z:Z\times\re^d\to Z$ is the projection, $\mathrm{p}_z(z,u)=z$. Consider the
set
$$\tld U_{n}=\left\{y\in \re^d: \inf_{u\in U}|y-u|<\frac{1}{n}\right\}\cap\left\{y:||y|-1|<\frac{1}{n}\right\}.$$
Naturally, $\tld U_n$ is open and $\bigcap_{n\geq 1}\tld U_n=U$. Finally, let $B(0,1/n)=\{y\in
\re^d: |y|<1/n\}$. The set $\Upsilon_n=\Gamma^{-1}(B(0,1/n))\cap(Z\times\tld U_n)$ is open and thus
its image by the projection $\mathrm{p}_z$ is a measurable set. Furthermore, we have
\begin{equation}
\begin{split}
\bigcap_{n\geq 1}\mathrm{p}_z(\Upsilon_n)=&\left\{z\in Z: \forall n\; \Gamma(z)^{-1}(B(0,1/n))\cap\tld
U_n\neq \emptyset\right\}\\
=&\left\{z\in Z: \forall n\;\exists_{\beta_n\in \Gamma(z)^{-1}(B(0,1/n))\cap \tld
U_n}\; |\Gamma(z)\beta_n|<\frac{1}{n}\right\}\\
=&\left\{z\in Z: \exists_{\beta\in \tld U}\; \Gamma(z)\beta=0\right\}=\mathrm{p}_z\left(\Gamma^{-1}(\{0\})\cap(Z\times\tld U)\right),
\end{split}
\end{equation}
where the last equalities follow by choosing a converging subsequence $\beta_{n_k}\to \beta$
and observing that $\beta\in Ker(\Gamma(z))\cap \tld U$. In consequence, $F^-(U)$ is an intersection of measurable sets and $F$ is measurable. As we work in metric spaces, $F$ is also weakly measurable
and an application of the Measurable Selection Theorem of Kuratowski and Ryll-Nardzewski (cf.~Wagner \cite[Thm 4.1]{WAG77}) completes the proof.\hfill$\square$

\bibliographystyle{abbrv}
\bibliography{complete_options}

\end{document}